\documentclass[aps,pre,preprint, showpacs, superscriptaddress]{revtex4}
\usepackage[dvips]{graphicx}
\usepackage{latexsym}
\usepackage{natbib}

\usepackage{color}
\usepackage{amssymb}
\usepackage{multirow}
\usepackage{bigstrut}

\begin{document}
\title {\bf Synchronization in interacting Scale Free Networks}

\author{M. F. Torres} \affiliation{Instituto de Investigaciones
  F\'isicas de Mar del Plata (IFIMAR)-Physics Department,
  Facultad de Ciencias Exactas y Naturales, Universidad Nacional de
  Mar del Plata-CONICET, Funes 3350, (7600) Mar del Plata, Argentina.}

\author{M. A. Di Muro} \affiliation{Instituto de Investigaciones
  F\'isicas de Mar del Plata (IFIMAR)-Physics Department,
  Facultad de Ciencias Exactas y Naturales, Universidad Nacional de
  Mar del Plata-CONICET, Funes 3350, (7600) Mar del Plata, Argentina.}
\author{C.~E.~La~Rocca} \affiliation{Instituto de Investigaciones
  F\'isicas de Mar del Plata (IFIMAR)-Physics Department,
  Facultad de Ciencias Exactas y Naturales, Universidad Nacional de
  Mar del Plata-CONICET, Funes 3350, (7600) Mar del Plata, Argentina.}
\author{L. A. Braunstein} \affiliation{Instituto de Investigaciones
  F\'isicas de Mar del Plata (IFIMAR)-Physics Department,
  Facultad de Ciencias Exactas y Naturales, Universidad Nacional de
  Mar del Plata-CONICET, Funes 3350, (7600) Mar del Plata, Argentina.}
\affiliation{Center for Polymer Studies, Boston University, Boston,
  Massachusetts 02215, USA}


\begin{abstract}
We study the fluctuations of the interface, in the steady state, of
the Surface Relaxation Model (SRM) in two scale-free interacting
networks where a fraction $q$ of nodes in both networks interact one
to one through external connections. We find that as $q$ increases the
fluctuations on both networks decrease and thus the synchronization 
reaches an improvement of nearly $40\%$ when $q=1$. The
decrease of the fluctuations on both networks is due mainly to the
diffusion through external connections which allows to reducing the
load in nodes by sending their excess mostly to low-degree
nodes, which we report have the lowest heights. This effect enhances
the matching of the heights of low-and high-degree nodes as $q$
increases reducing the fluctuations. This effect is almost independent
of the degree distribution of the networks which means that the
interconnection governs the behavior of the process over its
topology.
\end{abstract}

\maketitle
\section{Introduction}

In the last decades the study of complex networks has been growing
strongly due to the large number of systems that exhibit this type of
structures. A complex network is a set of nodes that are connected by
internal links and the most fundamental property that characterizes
its topology is the degree distribution $P(k)$, which represents the
probability that a node has $k$ neighbors or connectivity $k$. It has
been found that many real systems such as social, communication and
biological networks present a degree distribution given by $P(k)\sim
k^{-\lambda}$, where $\lambda$ is the exponent of the power law and
$k_{min} \le k \le k_{max}$, where $k_{min}$ and $k_{max}$ are the minimum and maximum degree of the
network. These kind of networks are called Scale Free (SF) and one of
its most important features is that are in general very heterogeneous,
i.e. most nodes of the network have a low connectivity while only a
few have a high connectivity (hubs). In recent years the study of
synchronization processes in isolate complex networks has been
increasing because of its importance in neurobiology
\cite{neuro1,neuro2,neuro3,neuro4,neuro5} and population dynamics
\cite{pop1,pop2}. A common theoretical approach to study
synchronization in complex networks is to map this process onto an
interface growth model by assigning to each node a scalar field $h_i$,
with $i=1, \cdots, N$, where $N$ is the size of the network. This
scalar field could represent, for example, the amount of load on a
node in the problem of distributed parallel computing on
processors. Without loss of generality we will relate the scalar field
to a set of heights on the interface. The most relevant magnitude that
characterizes the interface is $W(t)\equiv W$, which represents the
fluctuations of the scalar field around its mean value on the network,
given by 
\begin{equation}
W=\left\{\frac{1}{N}\sum_{i=1}^{N}(h_i-\langle h \rangle)^{2}\right\}^{1/2},
\end{equation}
where $\langle h \rangle=\frac{1}{N}\sum_{i=1}^Nh_i$ is the average of the scalar field over the
nodes at time $t$ and $\left\{\;\right\}$ is the average over different
network realizations. The roughness evolves in time until it saturates
at the steady state. In the saturation regime $W_s$ is a constant that
depends only on $\lambda$. In complex networks the synchronization
of the system is related to the roughness in the saturation regime
\cite{Korniss,Ana,Cristian,Cristian2,Cristian3,Korniss3,Zoltan1,Zoltan2,Zoltan3}.

One of the most simple and used models to study synchronization in
complex networks is the Surface Relaxation Model (SRM)
\cite{Korniss,Ana,Cristian,Cristian2,Cristian3}  which is a modiﬁcation of the classic
Family model largely used in euclidean interfaces \cite{Family}. In this model, at
each time step, a node $i$ is randomly chosen and the node with the
lowest height between the chosen node and all its neighbors evolves
increasing its height. It has been found that for isolated SF
networks, with $\lambda<3$, $W_S \sim \ln N$
\cite{Ana,Cristian,Debora} until a critical value $N=N^*$, after which
$W_s$ becomes independent of $N$, which means that the system becomes
scalable \cite{Korniss, Debora}.  Although it was an interesting
result, many real systems are not isolated but interacting with other
systems instead. This means that a process that develops in one
network can be affected by a process developing in another and
vice versa
\cite{Bul_01,jia_02,Gregagos,Gao_12,Gao_01,Val13,Bax_01,Bru_01,Brummitt_12,Lee_12,Gomez_13,Kim_13,Cozzo_12,Car_02,Kal_13,Zhen_PR,Son}. Such
is the case of epidemic models where the interaction between networks
make it very harmful for the healthy populations because the
interaction increases the theoretical risk of infection compared with
the same process in isolated networks
\cite{Dickison_12,Yag_13,Zuz_14,Sanz_14,Sahneh_14,Buono2}.

These interacting systems can be modeled as networks that interact
through external links that connect nodes that belongs to different
networks. Now the question is, does synchronization in interacting
networks performs worse or better than synchronization in isolated
networks?. In order to answer this question, in this letter we study
the synchronization of two SF networks with the same size $N$ that
interact through a fraction $q$ of nodes connected, one by one,
between them. The model used is the SRM model which we adapt to
interacting networks and study the effect of the interaction parameter
$q$ on the fluctuations in both networks.
\section{Model}

\begin{figure}
\begin{center}
\includegraphics[width=0.49\textwidth]{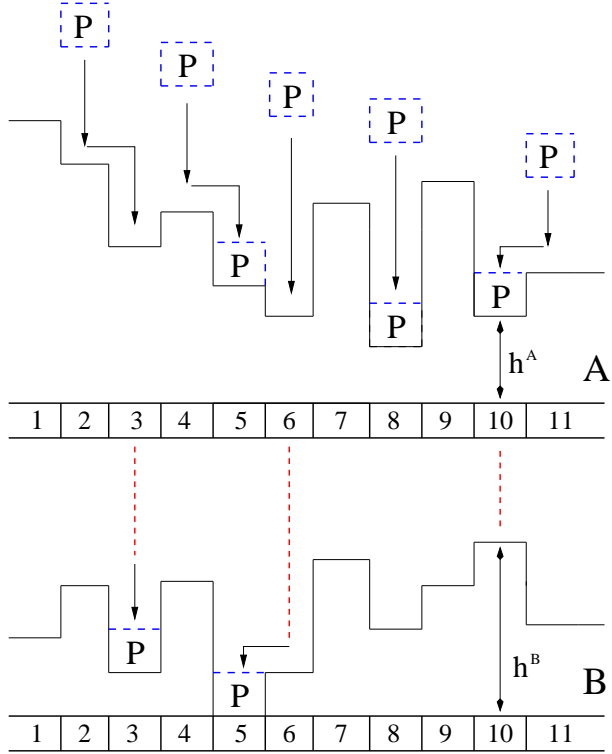}
\caption{Schematic of the rules of the model in a one dimensional
  Euclidean lattice. The particles $P$ are only dropped on the network
  $A$ in this scheme. The numbers represent different nodes in each
  network and the red dotted lines represent the external connections
  between nodes of different networks. In this case $N=11$ and
  $q=3/11$. The arrows indicate the path that the particles follow,
  which goes from the node where the particle was originally dropped,
  to the node where the particle gets finally deposited. The height of
  the nodes is measured from the upper line of the boxes that
  represent the numbers assigned to the nodes in each network.}
\label{fig.1}
\end{center}
\end{figure}

In our model, two uncorrelated SF networks, called $A$ and $B$, with
the same size $N$ and exponent $\lambda_{A}$ and $\lambda_{B}$
respectively are built using the Molloy-Reed algorithm \cite{Molloy}
disallowing self loops and multiple connections.  As we need a single
interface on each network, in order to ensure that we have a single
component we use $k_{min}=2$ \cite{Cohen}. Each node $i \in \alpha$,
with $\alpha=A,B$, has a connectivity $k_i^\alpha$ and we denote the
set of its neighbors by $v_i^{\alpha}$. In order to build the external
connections between the networks we connect by simplicity, the first
$q\;N$ nodes in $A$ one by one with the first $q\;N$ in $B$, where $q$
is the interaction parameter with $0< q \le 1$. Notice that
  if both networks are uncorrelated this procedure is the same as
  connecting a fraction q of nodes at random. We define the vector
$M$, where $M_i$, with $i=1,...,N$, is equal to $1$ if the node $i$ in
$A$ has an external connection with $i$ in $B$, and $M_i=0$
otherwise. In order to simplify the growth rules we choose random
initial conditions for the scalar field in the interval $[0,1]$,
hence, we avoid the cases in which different nodes have equal heights,
as we are only interested in the saturation regime on both networks
where the initial condition plays no role.

The evolution rules of the interface growth are given as follows:

\begin{enumerate}

\item{A network $\alpha$ (with $\alpha=A,B$) is chosen with
  probability $1/2$ and then a ``particle'', which represents the load,
  is dropped in a node $i$ selected randomly in $\alpha$.}

\item{The particle diffuses to the node $\epsilon$ that is the node
  with the lowest height between the node $i$ and its neighbors
  $v_i^{\alpha}$.}

\item{If $M_\epsilon=0$ or $M_\epsilon=1$ and $h_\epsilon \in \alpha
  <h_\epsilon \in \beta$ (with $\beta \neq \alpha$) the particle is
  deposited in $\epsilon \in \alpha$. Otherwise the particle diffuses
  to the network $\beta$ and is deposited in the node with the lowest
  height between $\epsilon$ and its neighbors $v_{\epsilon}^{\beta}$.}
\end{enumerate}

Thus if we denote $\ell \in \alpha$ as the node where the particle is finally deposited, then $h_\ell^{\alpha}=h_\ell^{\alpha} +1$. At each Monte-Carlo step the time is increased by $1/2N$. In Fig. \ref{fig.1} we show a schematic of the rules of the process for the case of a one-dimensional lattice.

\section{Results and Discussions}

\begin{figure}
\includegraphics[width=0.49\textwidth]{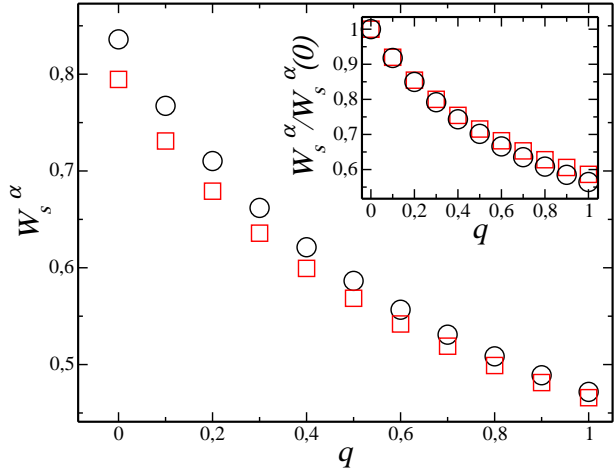}
\caption{$W_{s}^\alpha$ as a function of $q$ for $A$ $(\bigcirc)$ and $B$ $(\square)$ with $\lambda_{A}=2.6$ and $\lambda_{B}=3$. In the inset $W^{\alpha}_{S}/W^{\alpha}_{S}(q=0)$ as a function of $q$.}
\label{fig.2}
\end{figure}

We are interested in the behavior of the fluctuations in the steady
state of both networks with the system size above $N^*$ \cite{Debora},
value for which the system is scalable. For isolated SF networks this
regime for $\lambda<3$ is close to $N^* \approx 2 \times 10^{5}$
\cite{Debora}. We check that the nature of this regime is due to the
distribution of internal connectivities and that it is almost not
affected by the interaction parameter $q$. Thus in our research we use
$N=N_{A}=N_{B}= 3 \times 10^5$ in order to ensure that we are in the
scalable regime. We will show our results only for $\lambda_{A}=2.6$
and $\lambda_{B}=3$, because all the other combinations of the
exponents $\lambda$ in $2.5<\lambda\leq 3$ give qualitatively the same
results. We compute the fluctuations in the saturation regime of both networks
$W^{\alpha}_s=\sqrt{\frac{1}{N}\sum_{i=1}^N(h_i^\alpha - \langle
  h^\alpha\rangle)^2}$, with $\alpha=A,B$ and in Fig.~\ref{fig.2} we
show $W^{\alpha}_{S}$ as a function of $q$. It is clear that as the
interaction parameter $q$ increases, $W^\alpha_{S}$ decreases in both
networks, which implies that the synchronization improves. From the
plot we can also observe that as $q$ increases, the difference between
$W^{A}_{S}$ and $W^{B}_{S}$ becomes smaller, which means that the
synchronization in each network becomes mainly controlled by $q$ and
not by the internal degree distributions. In the inset of  Fig.~\ref{fig.2} we show $W^{\alpha}_{S}/W^{\alpha}_{S}(q=0)$ as a
function of $q$. It can be seen how is the rate of improvement in the
synchronization with the increment of $q$. For example, for $q=0.3$
the synchronization  enhances approximately $20\%$ and for $q=1$
around $40\%$.  It is worth pointing out that this rate decreases as $q$
  increases, and this means that the effect of the optimization gets less
  significant as the networks have more interconnections between them.

\begin{figure}
\includegraphics[width=0.49\textwidth]{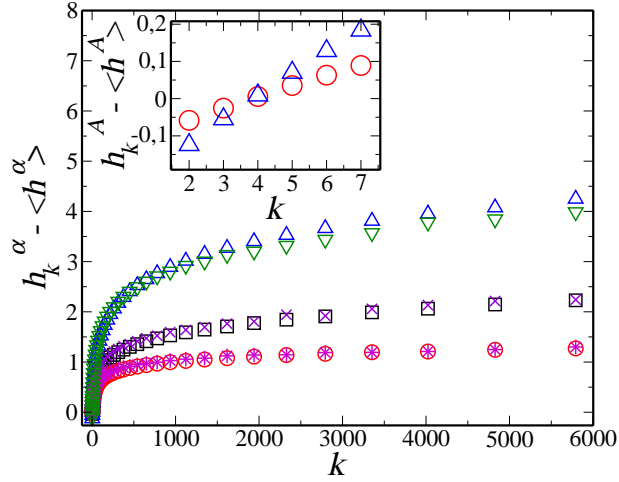}
\caption{$h_k^\alpha-\langle h^\alpha \rangle$ as a function of $k$
  for $q=0$ in $A$ ($\bigtriangleup$) and $B$ ($\bigtriangledown$),
  $q=0.5$ in $A$ ($\times$) and $B$ ($\Box$) and $q=1$ in $A$
  ($\bigcirc$) and $B$ ($\ast$). The inset is an amplification of
  $h_k^\alpha-\langle h^\alpha \rangle$ for the lowest values of $k$
  for $q=0$ in $A$ ($\bigtriangleup$) and $q=1$ in $A$ ($\bigcirc$)}
\label{fig.3}
\end{figure}

To understand the effect of the interaction parameter $q$ on the
optimization of the process we compute the difference between the
average height of nodes with degree $k$, denoted by $h_k^\alpha$ and
the mean value of the height $\langle h^{\alpha} \rangle$
as a function of $k$. In Fig. \ref{fig.3} we show $h_k^{\alpha}-
\langle h^{\alpha}\rangle$ as a function of $k$ for different values
of $q$. We can see that, for $q=0$ the heights of low connectivity
nodes, which are the majority in SF networks, are closer to the
average height of the network than the heights of high degree nodes,
which are above $\langle h^{\alpha} \rangle$. This means that hubs are
usually overloaded because all their neighbors send them their excess
of load, affecting negatively the synchronization of the
system. However, as the factor $q$ increases the height of the hubs
decreases, approaching to $\langle h^{\alpha} \rangle$ and becoming
independent of $k$ for $q=1$. In the inset of Fig. \ref{fig.3}, we can
see an amplification of the behavior of $h_k^{\alpha}$ for the nodes
with the lowest connectivity. We can see that as $q$ increases their
heights also approach to the average value. These results imply that
as we increase the factor $q$ hubs are no longer overloaded and
therefore their excess of load is now absorbed by low connectivity
nodes.

\begin{figure}
\includegraphics[width=0.49\textwidth]{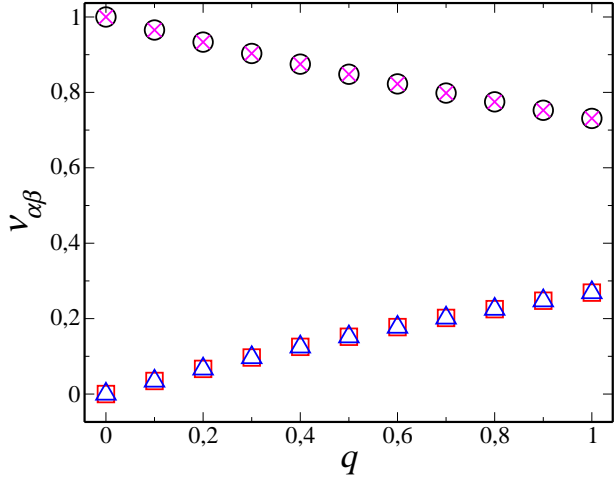}
\caption{Rates $\nu_{\alpha\beta}$ at which a particle spreads from the network $\alpha$ to the network $\beta$ as a function of $q$. With $\nu_{AA}$ ($\bigcirc$), $\nu_{AB}$ ($\Box$), $\nu_{BB}$   ($\times$) and $\nu_{BA}$ ($\bigtriangleup$). From the definition it is clear that $\nu_{AA}+\nu_{AB}=1$ and $\nu_{BA}+\nu_{BB}=1$.}
\label{fig.4}
\end{figure}

\begin{figure}
\includegraphics[width=0.49\textwidth]{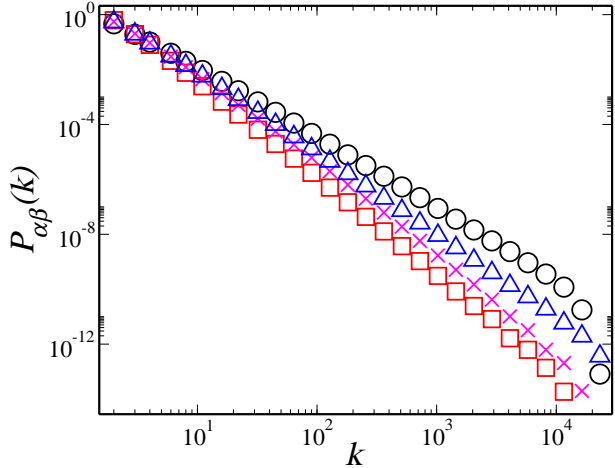}
\caption{Degree distributions $P_{\alpha\beta}(k)$ of the particles which get deposited on a node with connectivity $k$ on the network $\beta$ after being dropped on network $\alpha$ for $q=0.5$. With $P_{AA}$ ($\bigcirc$), $P_{AB}$ ($\Box$), $P_{BA}$ ($\times$) and $P_{BB}$ ($\bigtriangleup$).}
\label{fig.5}
\end{figure}

In order to understand the diffusion process between networks we want to know how frequently the load spreads from one network to another and how it gets distributed after diffusion. Hence we measure the rates $\nu_{\alpha\beta}$ at which a particle dropped in the network $\alpha$ gets finally deposited in the network $\beta$. In Fig. \ref{fig.4} we plot $\nu_{\alpha\beta}$ as a function of $q$. When $\alpha=\beta$ the rate always decreases with $q$, because the particles have more chances to cross to the other network. We can see that the rate is always much bigger when $\alpha=\beta$ than in the case of $\alpha\not=\beta$, which means that the particles tend to diffuse in the same network most of the time.  The fact that a small portion of load that crosses between networks is enough to enhance the synchronization is due to the fact that the dynamics of the SRM in both networks are coupled due to the interaction.  Another observation is that all the rates do almost not depend on the degree distribution of each network. The majority of nodes with $M_i\neq 0$ have an external connection with low connectivity nodes, due to the fact that these are selected at random and are the most common ones in SF networks. The fact that the amount of nodes with low connectivity does not change much with variations in $\lambda_\alpha$ for $2.5<\lambda_\alpha\leq 3$, and also because the diffusion between networks depends directly on the externally connected nodes whether a particle crosses to another network or remains in the same, make that the rates have almost no dependence on the exponents of the original degree distributions.

\begin{table}
\caption{Dispersion $\sigma_{\alpha \beta}$ of the distributions $P_{\alpha \beta}$ for different values of $q$} 
\label{TabDisp}
\begin{center}
\begin{tabular}{|c||c|c|c|c|}
\hline
\multirow{2}{0.6cm}{\centering $q$}& \multicolumn{4}{p{3.3cm}||}{\centering $\lambda_A=2.6$  $\lambda_B=3.0$} \bigstrut \\
\cline{2-5} &\multicolumn{1}{c|}{$\sigma_{AA}$} & \multicolumn{1}{c|}{$\sigma_{BB}$} & \multicolumn{1}{c|}{$\sigma_{AB}$} & \multicolumn{1}{c||}{$\sigma_{BA}$}  \bigstrut \\ \hline \hline
 0 & 22.1           & 7.5         &  $\textendash$       &\multicolumn{1}{c||}{$\textendash$  }         \bigstrut \\
 0.1 & 22.4           & 7.7         & 2.2        &\multicolumn{1}{c||}{3.3 }         \bigstrut \\
0.5 & 23.9           & 8.2         & 2.7        &\multicolumn{1}{c||}{4.5 }            \bigstrut \\
1.0 & 25.7           & 8.6         & 3.0       &\multicolumn{1}{c||}{5.7 }             \\ \hline 
\end{tabular}

\end{center}
\end{table}

How the load gets distributed after the diffusion process? In order
to answer this question we compute the probability
$P_{\alpha\beta}(k)$, defined as the probability that a particle
dropped in the network $\alpha$ gets deposited in a node with degree
$k$ in the network $\beta$. In Fig. \ref{fig.5} we plot
$P_{\alpha\beta}(k)$ with $\alpha,\beta=A,B$ for $q=0.5$. Also in
Table \ref{TabDisp} we report the dispersion
$\sigma_{\alpha\beta}$ of these distributions, which quantifies the
heterogeneity of the deposition process, for different values of $q$
and for the isolated networks. From the plot we can see that the
probabilities $P_{\alpha\alpha}(k)$ for $\alpha=A,B$ are very
heterogeneous and have a similar dispersion to the dispersion in the
isolated networks.  This means that when a particle stays on the
network where it was originally dropped finds an environment with a
similar heterogeneity than in the isolate SF network. Moreover the
distributions $P_{\alpha\beta}(k)$ for $\alpha \ne \beta$ are more
homogeneous than in the case $\alpha=\beta$ and thus have a lower
dispersion. This means that when a particle crosses from one network
to another it finds a more homogeneous neighborhood than in the case
in which stays in the same network, and as we will show below low-degree nodes are filled more often than high-degree nodes. Also we can
see that the dispersion, for all cases, slightly grows with $q$, and
this is due to the fact that hubs are less overloaded and can
participate more often in the diffusion process.

\begin{figure}
\includegraphics[width=0.49\textwidth]{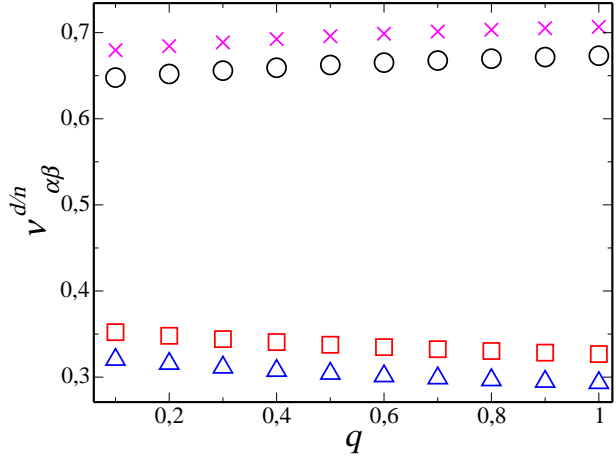}
\caption{Rates $\nu^{d}_{\alpha,\beta}$ and $\nu^{n}_{\alpha,\beta}$ at which a particle spreads from $\alpha$ to $\beta$ network and gets directly deposited or deposited on a neighboring node respectively. With $\nu^d_{AB}$ ($\bigcirc$), $\nu^n_{AB}$ ($\Box$), $\nu^d_{BA}$ ($\times$) and $\nu^n_{BA}$ ($\bigtriangleup$).}
\label{fig.6}
\end{figure}

\begin{figure}
\includegraphics[width=0.49\textwidth]{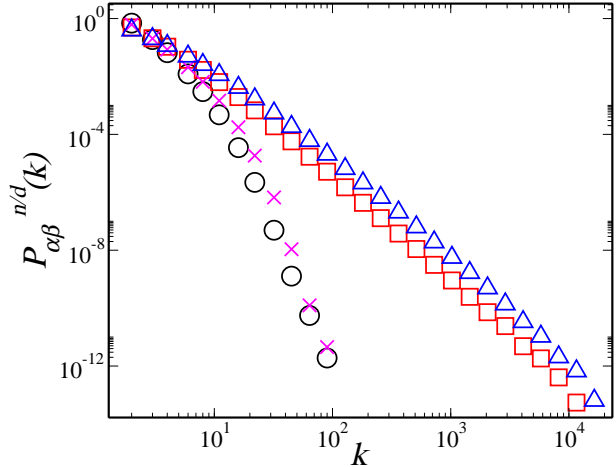}
\caption{Probabilities $P^{d}_{\alpha\beta}(k)$ and  $P^{n}_{\alpha\beta}(k)$, that after an external diffusion, the particle gets deposited directly or gets deposited on a neighboring node with degree $k$ respectively for $q=0.5$ . With $P^d_{AB}$ ($\bigcirc$), $P^n_{AB}$ ($\Box$), $P^d_{BA}$ ($\times$) and $P^n_{BA}$ ($\bigtriangleup$).}
\label{fig.7}
\end{figure}

We want to understand the reason that makes the load get rather deposited on low-connectivity nodes when it crosses to a different network. In order to explain this effect we study the diffusion process when the load crosses to the other network. When a particle spreads from one network to another, it can be directly deposited on a node connected by the external connection or it can be deposited in one of its neighbors.  To understand which of these two scenarios is more probable, we compute $\nu^{d}_{\alpha,\beta}$ and $\nu^{n}_{\alpha,\beta}$, which are the rates at which a particle that spreads from one network to another gets deposited directly ($d$) or in a neighboring ($n$) node respectively. Notice that $\nu^{d}_{\alpha,\beta}+\nu^{n}_{\alpha,\beta}=1$. In Fig. \ref{fig.6} we plot these rates as a function of $q$ and we can see that $\nu^{d}_{\alpha,\beta}$ is more important than $\nu^{n}_{\alpha,\beta}$, which means that for any $q$, most of the times the particles that cross from one network to another get directly deposited.

In order to explain the last observation, we define $P^{d}_{\alpha\beta}(k)$ and $P^{n}_{\alpha\beta}(k)$ as the probabilities that, after an external diffusion, the particle gets directly deposited or gets deposited on a neighboring node with degree $k$ respectively. In Fig. \ref{fig.7} we plot these probabilities for $q=0.5$. We can see that the probabilities $P^{d}_{\alpha\beta}(k)$ are more homogeneous than $P^{n}_{\alpha\beta}(k)$ and that nodes with low degree are the ones which receive the majority of the particles that are directly deposited. We can also see that the probabilities $P^{n}_{\alpha\beta}(k)$, which contemplate the scenario of neighboring deposition, have a wider spectrum, which agrees with the fact that in SF networks there are a few nodes with high degrees that receive the load by diffusion from their neighbors. This mechanism reduces the fluctuations, due to a matching of the heights of low-degree and high-degree nodes that is more efficient as $q$ increases. Finally the system is optimally synchronized for $q=1$.

\section{Conclusions}

We study the synchronization in two SF networks where the dynamic of
growth is ruled by a modified SRM model. We study the fluctuations in
the steady state of each network as a function of the interacting
parameter $q$ and we find that the synchronization of each
  network improves when the interconnection between them
  increases. This improvement in both networks is about a $40 \%$
  better than in isolated networks when $q=1$, which is an important
  value  regarding the decrease of the fluctuations. However, we show
that the rate of this improvement decrease with the number of
interconnections and this can be an useful result in future research
to determine if a larger interconnection between networks apart
from $q=1$, is worth it.

On the other hand, the improvement in the synchronization is due
mainly to the diffusion through external connections. We also found
that the majority of the particles that travel through the external
connections are directly deposited in nodes with low connectivity,
which are the majority in SF networks. We observe that in
average these nodes usually have a the lowest heights in a SF network. Then
when $q$ increases, the height of the nodes with low connectivity
increases  compared to the mean value and the difference with the
height of the hubs decreases. We also found various  distinctive
characteristics of the model, such as the fact that the synchronization
and the heights of the hubs become almost independent of the internal
degree distribution as $q$ increase, or the fact that the percentage
of particles that diffuse between networks only depends on $q$ and not
on the internal degree distribution of the  networks. This result
indicates that for high values of $q$ the behavior of the system is
ruled by the interconnection between networks and not by the topology
of the system.

\acknowledgments
MADM, CELR and LAB want to thanks UNMdP, FONCyT, Pict 0429/2013, CONICET, PIP 00443/2014 for financial support. MFT acknowledges CONICET, PIP 00629/2014 for financial support.

\end{document}